\documentclass[12pt,english]{article}
\usepackage[utf8]{inputenc}
\usepackage{color}
\usepackage{babel}
\usepackage{float}
\usepackage{amsmath}
\usepackage{amssymb}
\usepackage{graphicx}
\usepackage[a4paper]{geometry}
\usepackage[bookmarks=false,
 breaklinks=false,pdfborder={0 0 1},backref=section,colorlinks=true]
 {hyperref}
\hypersetup{
 linkcolor=blue,citecolor=blue,urlcolor=blue}

\makeatletter

\usepackage{setspace}
\usepackage{textcomp}

\makeatother

\begin{document}
\title{A toy model of a protein prototype reveals nontrivial ultrametricity
of the energy landscape}
\author{A.\,Kh.~Bikulov \\
 \textit{N.N. Semenov Federal Research Center for Chemical Physics,}
\\
 \textit{Russian Academy of Sciences,} \\
 \textit{Kosygin street 4, 117734 Moscow, Russia} \\
 e-mail:\thickspace{}\texttt{beecul@mail.ru} \\
 and \\
 A.\,P.~Zubarev \\
 \textit{Volga State Transport University,} \\
 \textit{Pervyi Bezymyannyi pereulok 18, Samara, 443066, Russia;}
\\
 \textit{Samara University,} \\
 \textit{Moskovskoe shosse 34, Samara, 443123 Russia} \\
 e-mail:\thickspace{}\texttt{apzubarev@mail.ru} }
\maketitle
\begin{abstract}
This work presents a theoretical and computational framework for investigating
the ultrametricity of the energy landscape in a model of a disordered
heteropolymer. This model is considered as the toy model of a protein
molecule, in which amino acid residues are modeled as material points.
In this model, pairwise interactions include universal repulsion,
the Lennard-Jones potential for hydrophobic residues, the Coulomb
potential with screening for charged residues, and an elastic potential
for bonds between neighboring residues. In studying this model, we
employ its analogy with spin glass models, which allows the application
of replica theory methods for the statistical description of the system.
However, in contrast to the standard approach to disordered systems,
averaging over disorder (over different sequences) is not performed.
Each sequence is analyzed independently, which is necessary for understanding
how a specific realization of disorder influences the formation of
hierarchical structure. In this case, the overlap between replicas
is defined as the Pearson correlation coefficient between vectors
of mean pairwise energies, which corresponds to comparing thermodynamic
averages in the spirit of spin glass theory. The paper presents the
results of a computational experiment conducted using a developed
algorithm on a GPU. The simulation used a chain of length $N=128$
residues, $K=50$ independent disorder realizations, $M=50$ replicas
for each sequence at temperature $T=1.0$ in arbitrary units, and
potentials with fixed parameters. It was found that for $90.0\%$
of sequences, the distance matrix between replicas contains more than
half of ultrametric triangles, and in the vast majority of such cases
($97.8\%$), nontrivial ultrametricity predominates, providing evidence
for the hierarchical organization of the energy landscape even within
this extremely simplified model. A repeated computational experiment
for selected sequences confirms the robustness of the observations:
$95.5\%$ of them again demonstrated ultrametricity, of which $97.7\%$
exhibited predominance of the nontrivial type of ultrametricity. The
obtained results support Frauenfelder's hypothesis about the ultrametricity
of proteins and open the way for a systematic study of ultrametric
properties in more realistic protein models.
\end{abstract}

\section{Introduction}

The concept of ultrametricity entered physics from mathematics, specifically
from number theory and functional analysis. In contrast to the Euclidean
metric, an ultrametric $d$ satisfies the so-called strong triangle
inequality: for any three points $x,y,z$, the condition $d(x,z)\le\max\{d(x,y),d(y,z)\}$
holds. Consequently, this modification of the axiomatic definition
of a metric leads to several important consequences: in an ultrametric
space, all triangles are isosceles with the base no longer than the
equal sides, and any two balls are either disjoint or one is completely
contained within the other. Such a structure is ideally suited for
describing hierarchical systems, where the states of a system can
be represented as a branching tree: the distance between two leaves
is determined by the level of their nearest common ancestor. It is
precisely this property that has made ultrametricity a key concept
in describing systems with complex energy landscape structure ---
from spin glasses to biological macromolecules.

The discovery of ultrametricity in condensed matter physics is inextricably
linked to the theory of spin glasses. Spin glasses are magnetic systems
with competing ferromagnetic and antiferromagnetic interactions, leading
to frustrations and a huge number of nearly degenerate ground states.
For a long time, the description of such systems remained an unsolvable
problem until 1979, when Parisi proposed a solution method for the
Sherrington--Kirkpatrick (SK) model by introducing an infinite number
of order parameters and the concept of spontaneous replica symmetry
breaking \cite{parisi1979,parisi1980}. Soon after Parisi's work,
it was realized that the hierarchy of order parameters has a direct
geometric interpretation \cite{mezard1984,rammal1986}. Namely, it
was shown that the space of pure states in a spin glass possesses
an ultrametric structure: overlaps (correlations) between states satisfy
the strong triangle inequality, and the states themselves can be organized
into a hierarchical tree. Subsequently, the development of various
methods for proving ultrametricity in spin glasses proceeded along
two directions: theoretical and computational. On the theoretical
side, the works of Talagrand played a key role, providing rigorous
mathematical foundations for Parisi's ideas \cite{talagrand2003,talagrand2006,talagrand2011}.
These works proved the Parisi formula for the free energy of the SK
model and laid the foundations for a rigorous approach to ultrametricity.
These ideas were further developed in the works of Panchenko, who
proved that ultrametricity is not merely a possible but a necessary
property of generalized SK models under certain symmetry conditions
\cite{panchenko2012,panchenko2013,panchenko2015}. Contemporary research
in this direction continues. For example, a recent work by Subag \cite{subag2024}
develops the Thouless--Anderson--Palmer (TAP) approach and shows
how the ultrametric tree of pure states can be naturally embedded
into the configuration space of spherical models, with points on this
tree approximately satisfying TAP equations for critical points.

Numerical studies of ultrametricity in spin glasses have been conducted
in parallel with theoretical ones. Monte Carlo simulations for the
three-dimensional Edwards--Anderson model, performed in a series
of works (see, e.g., \cite{ogielski1985,marinari1998,franz2000,contucci2007}),
confirmed the presence of hierarchical organization of states and
ultrametric properties of the overlap distribution. Later, with the
growth of computational power, works appeared that directly tested
the ultrametric inequality for configurations obtained from simulated
annealing (see, e.g., \cite{katzgraber2006}). These studies confirmed
that in the spin glass phase, overlap triangles are indeed isosceles,
with deviations from ultrametricity disappearing as temperature decreases
and system size increases. Modern numerical experiments on supercomputers
(see, e.g., \cite{mart=0000EDn2023}) continue to investigate the
limits of applicability of the ultrametricity hypothesis and its connection
with dynamical properties such as aging and slow relaxation. These
studies show that ultrametricity manifests not only in equilibrium
properties but also in nonequilibrium dynamics, where the system hierarchically
explores the valleys of the energy landscape. It is also interesting
to note that recent experimental work on photonic analogs of spin
glasses (random lasers) has confirmed the existence of ultrametric
structure in real physical systems \cite{ghofraniha2025}, demonstrating
the universality of this phenomenon.

The success of the ultrametric approach in spin glass physics has
influenced related fields, particularly biophysics. More than forty
years ago, Frauenfelder hypothesized the ultrametric nature of proteins.
The basis for this hypothesis was a series of experiments on carbon
monoxide (CO) binding to myoglobin \cite{frauenfelder1988,frauenfelder1991,frauenfelder1999}.
In these experiments, a myoglobin sample with bound CO was irradiated
with a laser pulse, breaking the bond, and then the kinetics of rebinding
were observed over a wide range of time scales --- from nanoseconds
to hundreds of seconds --- and over a broad temperature interval
(from 60 to 300 K). As a result of these experiments, it was discovered
that the kinetics of CO binding to myoglobin are strongly nonexponential,
with the shape of the curves depending on temperature in a complex
manner. Frauenfelder suggested that a protein could exist in many
different metastable substates organized hierarchically: substates
separated by small energy barriers are combined into clusters separated
by higher energy barriers, these clusters are combined into even larger
clusters, and so on. Such a hierarchical structure is naturally described
by an ultrametric tree, and protein dynamics can be described as a
random walk on this ultrametric tree \cite{avetisov2008}.

Attempts to prove the existence of ultrametricity in protein states
can be divided into three categories: theoretical modeling, analysis
of experimental data, and computational experiments in protein models.
The most consistent program for theoretical modeling of protein dynamics
within the ultrametric framework was developed in \cite{avetisov1999,avetisov2002,avetisov2013ultra,bikulov2021},
where the apparatus of $p$-adic analysis was proposed to describe
protein conformational dynamics. In this approach, the set of metastable
states (conformation space) of a protein is described by an ultrametric
space, which is taken to be the set of $p$-adic numbers (or some
subset thereof), and diffusion through this space is described by
$p$-adic pseudodifferential equations. In \cite{avetisov2002,avetisov2013ultra,bikulov2021},
it was shown that the same model of ultrametric diffusion with a reactive
sink describes the kinetics of CO binding to myoglobin over the entire
temperature range from 60 to 300 K. At high temperatures (200--300
K), the model predicts power-law kinetics at the initial stage, transitioning
to an exponential, while at low temperatures (below 180 K), it predicts
a purely power-law dependence, which exactly corresponds to experimental
observations. This result can be considered strong indirect evidence
in favor of the ultrametric organization of proteins, since no other
model has so far been able to uniformly describe the data over the
entire range of times and temperatures.

Direct experimental proof of protein ultrametricity has long remained
unattainable due to limitations in spatiotemporal resolution. However,
a recent breakthrough in super-resolution microscopy has opened new
perspectives. For example, work \cite{shaib2024} demonstrates the
possibility of visualizing individual protein shapes with approximately
1 nanometer resolution using one-step nanoscale expansion microscopy.
Although this work does not directly measure ultrametricity, it shows
that experimental observation of conformational landscapes at the
single-molecule level is becoming a reality, which may in the future
allow direct testing of ultrametric hypotheses.

In the field of computational experiments, there is also some work
providing evidence in favor of protein ultrametricity. For example,
the authors of \cite{velikson1993} conducted a computer study of
the conformational space of the poly-L-alanine peptide consisting
of 20 residues, using an all-atom model. Using molecular dynamics
and Monte Carlo methods, they generated an ensemble of conformations,
then performed minimization to search for local energy minima. The
main goal of the work was to test the hypothesis that the protein
energy landscape, like that of spin glasses, possesses the property
of ultrametricity. As a result, the authors found that the set of
low-energy conformations (alpha-helices) indeed exhibits an ultrametric
structure, indicating the presence of hierarchical organization among
states close in energy. There are also later works in the framework
of molecular dynamics that provide direct or indirect evidence in
favor of protein ultrametricity. In \cite{scalco2012}, the possibility
of applying an ultrametric approach to the analysis of protein folding
kinetics was investigated. The authors of this work used ergodic Markov
state models constructed from molecular dynamics trajectories and
showed that the free energy of the transition state between two nodes,
defined through the minimum cut, satisfies the axioms of ultrametricity.
This allowed the development of an automatic procedure for reducing
the model to a small number of macrostates, with the resulting rate
constants having clear physical meaning and agreeing with transition
state theory and Kramers' theory. This work provides strong computational
confirmation that the conformation space of proteins indeed possesses
an ultrametric topology. Another direction of computational research
is represented by \cite{mele2022}. Here, the authors applied an information-theoretic
approach based on the concepts of resolution and relevance to assess
the quality of various methods for clustering protein conformational
space. They investigated seven different hierarchical clustering strategies
on a molecular dynamics dataset for twelve structurally different
proteins. The results showed that the most physically adequate clusterings
are those that preserve the ultrametric structure of the low-dimensional
space. This study is interesting because it does not postulate ultrametricity
but rather discovers it as a natural property of optimal clusterings,
which serves as independent confirmation of Frauenfelder's hypothesis.

The present work lies at the intersection of theoretical modeling
and computational experiments in models of disordered polymers, which
we consider as models of protein prototypes. We consider a simple
model of a polymer chain with four types of residues, describe a theoretical
and computational scheme for investigating the ultrametricity of the
energy landscape of such a system, and conduct a computational experiment
to study ultrametricity in this model. This model represents an abstract
system consisting of a chain of point particles (by analogy with real
proteins we call them "residues"), sequentially connected by elastic
forces, and can be considered only as a highly simplified version
of some prototype of a polypeptide chain. Residues are classified
into four types: hydrophobic, hydrophilic positively charged, hydrophilic
negatively charged, and hydrophilic neutral. For all pairs of residues,
short-range repulsion exists. Hydrophobic residues interact pairwise
via the Lennard-Jones potential. Charged hydrophilic residues interact
via the Coulomb potential. Additionally, all residues that are neighbors
in the chain interact via an elastic potential.

It is important to emphasize that the proposed model is not a model
of any specific protein. By the term "protein prototype model" we
understand an abstract mathematical construction that imitates only
the most general features of protein macromolecules: a linear sequence
of different monomers (residues) and the presence of interactions
qualitatively resembling hydrophobic and electrostatic forces. This
model represents an extremely simplified, or "toy," system intended
for investigating fundamental statistical properties of the energy
landscape, rather than for quantitative description of real proteins.
All elements of the model (point particles, interaction potentials)
are chosen so that, on the one hand, key factors leading to frustration
and complex landscape (competition between attraction and repulsion,
diversity of monomer types) are preserved, while on the other hand,
computational complexity is minimized, making statistical analysis
possible using methods analogous to spin glass theory. This approach
fully corresponds to the spirit of "toy models" in condensed matter
physics, where the main goal is to identify qualitative patterns rather
than achieve chemical accuracy.

In constructing and analyzing the model, we use the analogy with spin
glass theory, which allows employing the apparatus of replica theory
for statistical analysis of the system. The analysis we apply is based
on the use of concepts of disorder (specific residue sequence), microstates
(specific spatial configurations), and macrostates (replicas as canonical
distributions of configurations). This enables statistical averaging
both over thermal fluctuations and over the ensemble of different
sequences. Nevertheless, the analysis of ultrametricity is performed
for each sequence separately, without averaging the overlap matrices
over disorders. To investigate ultrametricity, we use a computational
algorithm including the following stages: generation of an ensemble
of random residue sequences (disorder realizations); achievement of
thermal equilibrium for each sequence by the Monte Carlo method with
the Metropolis algorithm \cite{metropolis1953,newman1999}, simulated
annealing, and individual step-size adaptation for each replica; collection
of statistically independent configurations for constructing canonical
distributions (replicas); computation of the overlap matrix between
replicas based on comparison of pairwise energy distributions; verification
of ultrametricity of the obtained distance matrix separately for each
sequence.

It should be noted that when verifying ultrametricity, one should
distinguish between so-called "trivial ultrametricity" and "nontrivial
ultrametricity." Trivial ultrametricity occurs when distances between
any three points are equal. Such ultrametricity can arise in systems
without real hierarchy, for example, with random selection of points
in a high-dimensional space \cite{zubarev2014,zubarev2017}. Nontrivial
ultrametricity, on the contrary, is convincing evidence of a complex
hierarchical organization of the state space. It is precisely the
distinction between these cases that is of fundamental importance
for the correct interpretation of the results of our computational
experiment. Furthermore, it is also necessary to take into account
the fact that in models of non-ergodic systems, such as spin glasses
and proteins, different initial microstates during Monte Carlo simulation
may converge to the same macrostate (replica). Namely, upon random
initialization of system configurations, some configurations may over
time fall into the basin of attraction of the same macrostate, leading
to high overlap values (close to unity) between the corresponding
replicas. For a correct analysis of ultrametricity, one should consider
only those replicas that are guaranteed to belong to different macrostates.
Therefore, it is also necessary to use a procedure for selecting unique
replicas: if the overlap between two replicas exceeds a given threshold,
they are considered to belong to the same macrostate, and one of them
is excluded from further consideration. Thus, the analysis of ultrametricity
is performed only on the set of replicas corresponding to different
macrostates, which allows correctly identifying the hierarchical structure,
if it exists.

It is important to emphasize the conceptual difference between the
approach adopted in this work and the classical approach to disordered
polymers. From a formal point of view, our model with a fixed but
random residue sequence is a typical example of a disordered heteropolymer.
In condensed matter physics, the standard procedure for such systems
is averaging all observables over the ensemble of disorder realizations
(sequences), which allows obtaining a description of the typical behavior
of the system. However, in the context of protein research, such an
approach is not adequate. Each real protein is a specific sequence,
and its unique properties, including the structure of its energy landscape,
are determined precisely by this specificity. Therefore, in contrast
to the classical theory of disordered polymers, in this work we fundamentally
refrain from averaging over disorder. All stages of analysis ---
from constructing overlap matrices to classifying types of ultrametricity
--- are performed for each sequence (for each disorder realization)
independently. This allows not only estimating the fraction of sequences
demonstrating hierarchical organization but also, in perspective (when
transitioning to more realistic models), studying which specific features
of the amino acid sequence encode the complexity of the energy landscape,
which directly brings us closer to understanding the principles of
folding and functioning of real proteins.

The work is organized as follows. Section 2 presents a complete description
of our proposed toy model of a protein prototype. Section 3 describes
in detail the algorithm for investigating ultrametricity, including
all computational stages and statistical procedures. Section 4 presents
the simulation results, revealing nontrivial ultrametricity in the
considered model. The final section presents concluding remarks and
discussion of prospects for further research.

\section{Description of the toy model of a protein prototype for studying
ultrametricity}

In our model, a protein prototype is considered as a linear chain
consisting of $N$ ordered material points, in which neighboring points
are connected by linear elastic forces. By analogy with proteins,
we will also call these material points amino acid residues. From
a formal point of view, such a system represents a classic example
of a disordered heteropolymer: a linear chain consisting of monomers
(residues) of several types, whose sequence along the chain is fixed
but random. In real proteins, residues differ in their chemical properties,
which determines the specificity of their interactions. In our model,
we use a simplified classification including four types of residues:
hydrophobic, hydrophilic positively charged, hydrophilic negatively
charged, and hydrophilic neutral. This division allows accounting
not only for the hydrophobic effect but also for electrostatic interactions,
which play an important role in stabilizing protein structures and
can generate additional frustrations.

Formally, the residue sequence is defined as a vector $\sigma=(\sigma_{1},\sigma_{2},\dots,\sigma_{N})$,
where each element $\sigma_{i}$ encodes the residue type. We introduce
two characteristics for each residue: charge $q_{i}\in\{-1,0,+1\}$
and hydrophobicity indicator $h_{i}\in\{0,1\}$. Hydrophobic residues
($h_{i}=1$, $q_{i}=0$) interact via the Lennard-Jones potential.
Charged residues ($q_{i}=\pm1$, $h_{i}=0$) participate in Coulomb
interactions. Neutral hydrophilic residues ($q_{i}=0$, $h_{i}=0$)
have no specific interactions, except for universal repulsion.

The ensemble of different sequences ${\sigma^{(k)}}_{k=1}^{K}$ constitutes
the realization of disorder in the system, characteristic of disordered
heteropolymers. However, as already noted in the Introduction, in
contrast to the standard statistical mechanics of disordered systems,
where averaging over this ensemble is performed, in this work each
sequence $k$ is considered as unique and analyzed separately. We
consider $K$ independent sequences, each generated randomly: each
residue independently receives its corresponding type with given probabilities
$p_{\text{hp}}$ (hydrophobic), $p_{+}$ (positive), $p_{-}$ (negative),
and $p_{\text{neut}}=1-p_{\text{hp}}-p_{+}-p_{-}$ (neutral). All
sequences are considered equally probable.

A microstate of the system is considered to be a specific spatial
configuration of the sequence. The configuration is specified by a
set of residue coordinates $R=\{r_{1},r_{2},\dots,r_{N}\}$, $r_{i}\in\mathbb{R}^{3}$.
The energy function of the model has the form of a sum of several
contributions:

\begin{equation}
E(R,\sigma)=\sum_{i<j}U_{\text{pair}}\left(r_{ij},q_{i},q_{j},h_{i},h_{j}\right)+\sum_{i=1}^{N-1}U_{\text{bond}}\left(r_{i,i+1}\right),\label{E}
\end{equation}
where $r_{ij}=\left|\vec{r}_{i}-\vec{r}_{j}\right|$, and $q_{i}$,
$h_{i}$ are determined by the residue type $\sigma_{i}$. The pairwise
potential $U_{\text{pair}}$ includes universal short-range repulsion,
Lennard-Jones attraction potential, Coulomb potential, and elastic
potential. The universal short-range repulsion acts for all pairs
of residues regardless of their type:

\[
U_{\text{rep}}(r_{ij})=\begin{cases}
\varepsilon_{\text{rep}}\left(\dfrac{\sigma_{\text{rep}}}{r_{ij}}\right)^{12}, & \text{if }r_{ij}<\sigma_{\text{rep}},\\
0, & \text{if }r_{ij}\geq\sigma_{\text{rep}}.
\end{cases}
\]
The Lennard-Jones attraction potential is taken into account only
between hydrophobic residues ($h_{i}=h_{j}=1$) that are not nearest
neighbors along the chain ($|i-j|>1$), at distances not exceeding
the cutoff radius $r_{\text{cut}}$:

\begin{equation}
U_{\text{LJ}}(r_{ij})=4\varepsilon_{\text{LJ}}\left[\left(\frac{\sigma_{\text{LJ}}}{r_{ij}}\right)^{12}-\left(\frac{\sigma_{\text{LJ}}}{r_{ij}}\right)^{6}\right].\label{L_D}
\end{equation}
The Coulomb potential is taken into account for all pairs of charged
residues ($q_{i}\neq0$, $q_{j}\neq0$), also excluding nearest neighbors.
To avoid divergence at small distances, a minimum distance $r_{\text{min}}$
is introduced, and Debye screening is also taken into account:

\begin{equation}
U_{\text{coul}}(r_{ij})=\left\{ \begin{array}{c}
k_{\text{coul}}\,\dfrac{q_{i}q_{j}}{r_{ij}}\,\exp\left(-\frac{r_{ij}}{\lambda_{\text{Debye}}}\right),\;r_{ij}\ge r_{\text{min}},\\
k_{\text{coul}}\,\dfrac{q_{i}q_{j}}{r_{\text{min}}},\;r_{ij}<r_{\text{min}}.
\end{array}\right.\label{Coul}
\end{equation}
Here $k_{\text{coul}}$ is a constant determining the strength of
electrostatics, $\lambda_{\text{Debye}}$ is the screening length.
Thus, the total pairwise potential for a pair $(i,j)$ has the form:

\[
U_{\text{pair}}(r_{ij},q_{i},q_{j},h_{i},h_{j})=U_{\text{rep}}(r_{ij})+\delta_{h_{i},1}\delta_{h_{j},1}\delta_{|i-j|>1}U_{\text{LJ}}(r_{ij})+\delta_{q_{i}\neq0}\delta_{q_{j}\neq0}\delta_{|i-j|>1}U_{\text{coul}}(r_{ij}),
\]
where we use the notation $\delta_{A}=1$ if condition $A$ is satisfied
and $\delta_{A}=0$ if condition $A$ is not satisfied. The elastic
potential $U_{\text{bond}}$ acts between neighboring residues of
the chain and ensures its integrity:

\begin{equation}
U_{\text{bond}}(r_{i,j})=\frac{1}{2}k_{\text{bond}}(r_{i,j}-r_{\text{bond}})^{2}\delta_{|i-j|=1},\label{Bond}
\end{equation}
where $r_{\text{bond}}$ is the equilibrium bond length, $k_{\text{bond}}$
is the bond stiffness.

To characterize the state of the system, we use the concept of a pairwise
energy vector. For each configuration $R$, a vector $u\in\mathbb{R}^{N_{p}}$
is computed, where $N_{p}=\dfrac{N(N-1)}{2}$ is the number of unique
residue pairs ($i<j$). The components of this vector are the values
of the total pairwise potential $U_{\text{pair}}$ for all pairs,
including contributions from repulsion, Lennard-Jones, Coulomb, and
elastic interactions. Thus, each configuration is mapped to a point
$u$ in the pairwise energy space of dimension $N_{p}$, and this
point carries information about all interactions in the system.

A macrostate of the system is described by the canonical Gibbs distribution.
For each fixed residue sequence $\sigma^{(k)}$ (where $k=1,\dots,K$
is the index of the amino acid sequence (disorder realization)), the
probability of finding the system in configuration $R$ at temperature
$T$ is determined by the Boltzmann distribution:
\[
P(R|\sigma^{(k)})=Z^{-1}\left(\sigma^{(k)}\right)\exp\left(-\beta E\left((R,\sigma^{(k)}\right)\right),
\]
where $\beta=\dfrac{1}{k_{B}T}$ is the inverse temperature, $E\left((R,\sigma^{(k)}\right)$
is the energy of configuration $R$ for sequence $\sigma^{(k)}$,
$Z(\sigma^{(k)})=\int dR\exp\left(-\beta E\left((R,\sigma^{(k)}\right)\right)$
is the partition function. In the context of replica theory \cite{parisi1979,parisi1980,mezard1984},
such macrostates will be called replicas. Let $m=1,\dots,M$ be the
replica index, $s=1,\dots,S$ be the index of a statistically independent
realization of a microstate within a replica, $R^{(k,m,s)}$ be the
configuration corresponding to sequence $k$, replica $m$, and realization
$s$. Then each replica $m$ for sequence $k$ is represented by a
set of $S$ statistically independent configurations $\{R^{(k,m,s)}\}_{s=1}^{S}$,
sampled from the Boltzmann distribution. For a fixed amino acid sequence
$k$, the set of all possible configurations $R^{(k,m,s)}$ forms
the phase space of the system.

The central point of our methodology is the definition of overlap
between two replicas. As is known, in spin glass theory, the overlap
between two replicas $\alpha$ and $\beta$ is defined through thermodynamic
averages:

\[
q^{\alpha\beta}=\frac{1}{N}\sum_{i=1}^{N}\langle s_{i}\rangle_{\alpha}\langle s_{i}\rangle_{\beta},
\]
where $\langle s_{i}\rangle_{\alpha}$ is the average value of spin
$i$ over all configurations in replica $\alpha$, and $N$ is the
total number of spins. This definition is fundamentally important:
it compares not specific microstates but characteristics of distributions
--- the average magnetizations. In this case, the overlap $q^{\alpha\beta}$
takes values in the interval $[-1,1]$, which follows from the fact
that the average magnetizations of individual spins $\langle s_{i}\rangle_{\alpha}$
also lie in this interval. Drawing a direct analogy for our model,
we introduce the following definition of overlap between replicas
$m$ and $n$ for a fixed sequence $k$. For each replica, we compute
the vector of mean pairwise energies $\bar{u}^{(k,m)}\in\mathbb{R}^{N_{p}}$,
whose components are the thermodynamic averages of the total pairwise
energy for each pair of residues:

\[
\bar{u}_{p}^{(k,m)}=\langle u_{p}\rangle_{k,m}=\frac{1}{S}\sum_{s=1}^{S}u_{p}^{(k,m,s)},
\]
where $p=1,\dots,N_{p}$ enumerates all unique pairs $(i,j)$ with
$i<j$.

To obtain a definition of overlap that (i) takes values in $[-1,1]$
as in spin glass theory, (ii) is invariant under linear transformations
of the energy scale, and (iii) has a clear statistical interpretation
as a measure of linear correlation between distributions, we use the
Pearson correlation  coefficient between the vectors of mean pairwise
energies:

\begin{equation}
q_{mn}^{(k)}=\frac{{\displaystyle \sum_{p=1}^{N_{p}}\left(\bar{u}_{p}^{(k,m)}-\bar{U}^{(k,m)}\right)\left(\bar{u}_{p}^{(k,n)}-\bar{U}^{(k,n)}\right)}}{\sqrt{{\displaystyle \sum_{p=1}^{N_{p}}\left(\bar{u}_{p}^{(k,m)}-\bar{U}^{(k,m)}\right)^{2}}}\;\sqrt{{\displaystyle \sum_{p=1}^{N_{p}}\left(\bar{u}_{p}^{(k,n)}-\bar{U}^{(k,n)}\right)^{2}}}},\;\bar{U}^{(k,m)}=\frac{1}{N_{p}}\sum_{p=1}^{N_{p}}\bar{u}_{p}^{(k,m)}.\label{corr}
\end{equation}
The physical interpretation of this definition is as follows. If in
two replicas the system spends time in configurations where the average
interaction energies for each pair of residues are close (up to a
linear transformation), then such replicas are considered similar,
and their overlap is close to unity. If the distributions differ significantly
(for example, in one replica configurations with strong attraction
of hydrophobic residues dominate, while in another configurations
with predominance of repulsion dominate), then the overlap will be
small or even negative. The use of centering and normalization eliminates
the influence of absolute energy values, leaving only information
about the shape of the distributions.

After computing all pairwise overlaps for sequence $k$, an overlap
matrix $q_{mn}^{(k)}$ of dimension $M\times M$ is formed. The distance
matrix is defined through the correlation coefficient as $D_{mn}^{(k)}=1-q_{mn}^{(k)}$.
This definition guarantees that $D_{mn}^{(k)}\in[0,2]$, with $D_{mn}^{(k)}=0$
corresponding to identical distributions, and $D_{mn}^{(k)}=2$ corresponding
to the maximum possible difference (anticorrelation). It is precisely
these matrices, individual for each sequence, that are subjected to
further analysis for ultrametricity. Averaging of the overlap matrices
over the ensemble of sequences is not performed, as it would destroy
information about the different types of energy landscapes inherent
in different disorder realizations.

\section{Algorithm for investigating the ultrametricity of the energy landscape}

The algorithm we use represents a computational scheme for testing
the presence of ultrametricity in the energy landscape of the protein
prototype model described in the previous section. The algorithm consists
of a sequence of stages, each implementing a specific physical or
mathematical procedure. An important feature of the algorithm is its
parallelizability: computations can be performed simultaneously for
different disorder realizations (residue sequences) and for different
replicas (independent thermal ensembles). This allows efficient use
of modern computational resources, including graphics processing units.

Let us describe this algorithm stage by stage.

Stage 1. Generation of an ensemble of random residue sequences. At
this stage, parameters are set: number of sequences $K$, chain length
$N$, probabilities of three residue types $p_{\text{hp}}$ (hydrophobic),
$p_{+}$ (hydrophilic positively charged), $p_{-}$ (hydrophilic negatively
charged); the probability of the fourth type (hydrophilic neutral)
is computed automatically. Each sequence is generated independently
using a fixed random number generator seed, which ensures reproducibility
of results. For different sequences, different independent random
streams are used, effectively implemented by the generator, but the
overall seed is fixed. This stage corresponds to creating an ensemble
of different sequences with a given composition but random order of
residues.

Stage 2. Initialization of configurations for each sequence and each
replica. At this stage, for each sequence $\sigma^{(k)}$ ($k=1,\dots,K$),
$M$ independent replicas are created, and the initial coordinates
of residues are initialized. We initialize the initial coordinates
of all residues randomly on the surface of a sphere of sufficiently
small radius. When generating initial configurations, a fixed seed
is also used for reproducibility. Within a single residue sequence,
the initial coordinates of residues are chosen to be the same. This
is done to reduce the probability of the system converging to macrostates
that differ significantly in their spatial distribution.

Stage 3. Achievement of thermal equilibrium for each replica. At this
stage, the Monte Carlo method with the Metropolis algorithm is used.
At each step of the algorithm, a new configuration $R'=R+\delta R$
is proposed, where $\delta R$ is a random displacement with a Gaussian
distribution. The displacement amplitude is controlled and adapted
individually for each replica. If the acceptance rate for a given
replica falls below the target value (by default, this value is $0.3$),
the step size is decreased; if it is higher, the step size is increased.
Such individual adaptation allows each replica to efficiently explore
its region of phase space independently of others. To prevent the
system from getting stuck in local minima, simulated annealing is
used. The procedure starts with a low value of inverse temperature
$\beta=\dfrac{1}{k_{B}T}$ (which corresponds to a high temperature
$T$) and gradually increases $\beta$ to the target value according
to a linear law: $\beta_{\text{current}}=\beta_{\text{initial}}+(\beta_{\text{final}}-\beta_{\text{initial}})\cdot\dfrac{\text{step}}{\text{total\_steps}}$.
The energy change $\Delta E=E(R',\sigma)-E(R,\sigma)$ is computed
using formula (\ref{E}), including all energy contributions. The
transition to the new state is accepted with probability $P_{\text{accept}}=\min\left(1,\exp\left(-\beta\Delta E\right)\right)$.
The procedure is repeated for $N_{\text{eq}}$ steps until thermal
equilibrium is reached. The criterion for equilibrium is stabilization
of the average energy.

Stage 4. Collection of statistically independent configurations for
each replica. At this stage, after reaching equilibrium, $N_{\text{prod}}$
Monte Carlo steps are performed, during which the current configuration
is saved every $\tau$ steps. The parameter $\tau$ is chosen to be
larger than the correlation time of the system, which guarantees the
statistical independence of the saved configurations. As a result,
for each replica, a set of $S=\dfrac{N_{\text{prod}}}{\tau}$ configurations
$\{R^{(k,m,s)}\}_{s=1}^{S}$ is obtained, representing a sample from
the canonical distribution. These configurations characterize the
macrostate (replica) as a statistical ensemble.

Stage 5. Computation of the overlap matrix between replicas. At this
stage, for each sequence $k$ independently, the overlap matrix $q_{mn}^{(k)}$
of dimension $M\times M$ is computed according to definition (\ref{corr}).
Next, for each sequence $k$, the distance matrix $D_{mn}^{(k)}=1-q_{mn}^{(k)}$
is determined. At this stage, a key difference of our approach from
the standard analysis of disordered heteropolymers is realized. The
overlap matrix $q_{mn}^{(k)}$ and, consequently, the distance matrix
$D_{mn}^{(k)}$ are computed for each sequence $k$ independently.
No averaging of these matrices over the ensemble of sequences (i.e.,
over the disorder index $k$) is performed. All subsequent steps of
analysis (selection of unique replicas, verification of ultrametricity)
are also performed individually for each disorder realization. Additionally,
at this stage, the Edwards--Anderson parameter $q_{\text{EA}}^{(k)}$
is computed as the average of the off-diagonal elements of the overlap
matrix:
\[
q_{\text{EA}}^{(k)}=\frac{2}{M(M-1)}\sum_{m<n}q_{mn}^{(k)},
\]
which serves as an indicator of the spin glass phase.

Stage 6. Verification of ultrametricity for each sequence. Before
analyzing the ultrametric structure, it is necessary to ensure that
the considered replicas indeed represent different macrostates. In
non-ergodic systems, such as spin glasses and proteins, different
initial microstates may converge to the same macrostate (replica).
If the system is non-ergodic, then there exist many different macrostates.
Upon random initialization, some replicas may fall into the basin
of attraction of the same macrostate, leading to overlap values close
to unity. To correctly identify hierarchical organization, it is necessary
to exclude such "duplicate" replicas and perform the analysis only
on the set of replicas corresponding to different macrostates. In
the proposed algorithm, we introduce a procedure for selecting unique
replicas based on the overlap matrix. For a fixed sequence $k$, all
pairs of replicas $(m,n)$ are considered. If the condition $q_{mn}^{(k)}>1-\varepsilon_{\text{rep}}$
is satisfied, where $\varepsilon_{\text{rep}}$ is a user-defined
threshold (in this work, $\varepsilon_{\text{rep}}=0.2$ is adopted),
then such replicas are considered to belong to the same macrostate.
From each group of equivalent replicas, only one is retained; the
others are excluded from further consideration. As a result, for sequence
$k$, a reduced set of $M'_{k}\le M$ unique replicas is formed. The
corresponding distance matrix $D^{(k)}$ is truncated to dimension
$M'_{k}\times M'_{k}$.

After selecting unique replicas, the distance matrix $D^{(k)}$ is
analyzed according to the following algorithm. For each sequence $k$,
all triangles formed by triples of replicas $(i,j,l)$ from $M'_{k}$
are considered. For each triangle, the three distances $D_{ij}$,
$D_{i\ell}$, $D_{j\ell}$ are computed and ordered in increasing
order: $D_{\min}\le D_{\text{mid}}\le D_{\max}$. Then, using given
accuracy parameters $\epsilon$ and $\delta$ with $\delta>\varepsilon$,
the triangle is classified as follows:

1. If ${\displaystyle \frac{D_{\max}-D_{\min}}{D_{\min}}<\epsilon}$,
the triangle is considered trivially ultrametric.

2. If ${\displaystyle \frac{D_{\max}-D_{\text{mid}}}{D_{\text{mid}}}<\epsilon}$
and ${\displaystyle \frac{D_{\text{mid}}-D_{\min}}{D_{\text{mid}}}\ge\delta}$,
the triangle is considered nontrivially ultrametric.

3. Otherwise, the triangle is classified as non-ultrametric.

Next, for each sequence $k$, the fraction of trivially ultrametric
triangles $f_{\text{triv}}^{(k)}$, the fraction of nontrivially ultrametric
triangles $f_{\text{nontriv}}^{(k)}$, the total fraction of ultrametric
triangles $f_{\text{ultra}}^{(k)}=f_{\text{triv}}^{(k)}+f_{\text{nontriv}}^{(k)}$,
and the fraction of non-ultrametric triangles $f_{\text{non}}^{(k)}=1-f_{\text{ultra}}^{(k)}$
are computed. A sequence is considered to demonstrate ultrametricity
if $f_{\text{ultra}}^{(k)}>0.5$. In this case, if $f_{\text{nontriv}}^{(k)}>f_{\text{triv}}^{(k)}$,
the ultrametricity is classified as nontrivial; otherwise, it is classified
as trivial.

Stage 7. Classification of ultrametricity type and aggregation of
results. At this stage, based on the classification of individual
sequences, summary statistics over the ensemble are formed. Specifically,
the total number and fraction of sequences for which ultrametricity
was detected, the number and fraction of sequences with predominance
of nontrivial and trivial ultrametricity, and the average fractions
of trivial, nontrivial, total ultrametric, and non-ultrametric triangles,
as well as their standard deviations, are determined.

Stage 8. Repeated simulation. After completing Stage 7, those sequences
for which nontrivial ultrametricity was detected are selected. Then,
for the selected sequences, a repeated simulation (Stages 1--7) is
performed with the same model and simulation parameters but with a
changed random number generator seed. This ensures independence of
thermal fluctuations and initial conditions from the first stage.

The described algorithm completely excludes the operation of averaging
overlap matrices over the ensemble of sequences. Each sequence is
considered as an independent disorder realization with its own unique
energy landscape. This allows correctly estimating what fraction of
sequences demonstrate nontrivial ultrametricity and identifying sequence
parameters favorable for the formation of hierarchical structure.

\section{Simulation results}

This section presents the results of a computational experiment conducted
in accordance with the algorithm described above. The simulation was
performed on an NVIDIA Tesla P100 graphics processor using a developed
program in the Python language with the PyTorch library. Below are
the main parameters of the model and simulation, justification for
the choice of initialization method, detailed numerical results of
the first and second stages, and their analysis.

The simulation parameters in the numerical experiment were chosen
as follows. The chain length is $N=128$ residues, the number of independent
disorder realizations (number of sequences) is $K=50$, the number
of replicas for each sequence is $M=50$, the inverse temperature
is $\beta=1.0$ (temperature $T=1.0$ in units of $k_{B}=1$). To
improve convergence, simulated annealing was used with an initial
temperature $T_{initial}=10.0$. The number of steps to reach equilibrium
was $n_{equil}=100\,000$, the number of production steps at the statistics
collection stage was $n_{prod}=40\,000$, the configuration saving
interval was $\tau=100$ (thus, each replica is represented by $S=400$
independent configurations). The initial step size in the Metropolis
algorithm was chosen as $0.15$ with individual adaptation for each
replica to maintain an acceptance rate around $0.3$. Ultrametricity
verification parameters: $\varepsilon=0.05$, $\delta=0.1$, threshold
for identical replicas $\varepsilon_{rep}=0.2$. Probabilities of
residue types: $p_{hp}=0.4$ (hydrophobic), $p_{+}=0.25$ (positively
charged), $p_{-}=0.25$ (negatively charged), $p_{neut}=0.1$ (neutral).
Potential parameters: $\sigma_{rep}=1.0$, $\varepsilon_{rep}=1.0$
(repulsion); $\sigma_{LJ}=1.0$, $\varepsilon_{LJ}=5.0$, $r_{cut}=4.0$
(Lennard--Jones); $r_{bond}=1.0$, $k_{bond}=10.0$ (elastic bonds);
$k_{coul}=1.0$, $\lambda_{Debye}=2.0$, $r_{min}=0.5$ (Coulomb interaction
with screening).

It is important to note that the choice of specific values for the
accuracy parameters $\varepsilon=0.05$ and $\delta=0.1$ is purely
heuristic.These values were determined empirically. They provide a
reasonable balance between the strictness of the ultrametricity criterion
and the sensitivity of the method to the presence of hierarchical
structure, given the inevitable statistical fluctuations inherent
in computational experiments. Too small values of $\varepsilon$ and
$\delta$ would lead to the vast majority of triangles being classified
as non-ultrametric due to statistical noise, even in the presence
of clear hierarchical organization. Too large values, on the contrary,
could classify as ultrametric triangles that do not satisfy the strict
triangle inequality. The chosen values, based on preliminary test
calculations, allowed achieving this balance and, as will be shown
below, revealed a clear picture of the predominance of nontrivial
ultrametricity. Nevertheless, the question of the influence of the
values of $\varepsilon$ and $\delta$ on the final results and the
search for their optimal values remains a subject of separate investigation.

A remark should also be made regarding the choice of interaction potential
parameters listed above. These parameters are not universal and were
not extracted from any experimental data, which is in principle impossible
for such a coarse model. On the contrary, they were selected during
preliminary computational experiments with the aim of creating conditions
in which the ultrametric properties of the energy landscape could
manifest most clearly. In other words, we purposefully attempted to
find parameter values at which the system demonstrates a sufficiently
high level of frustration (to avoid trivial behavior) and retains
the ability to form a hierarchical structure of states. The criterion
for such selection was the maximization of the fraction of sequences
with nontrivial ultrametricity in several test runs. It is also important
to emphasize that a complete scan of the multidimensional parameter
space (interaction energies, screening length, bond stiffness, etc.)
is an extremely resource-intensive task, beyond the scope of this
study. Therefore, the values presented in the work, although allowing
successful verification of the presence of ultrametricity in the model,
nevertheless do not claim to be optimal or unique.

When conducting the simulation, we used initialization of the initial
coordinates of all residues randomly on the surface of a sphere of
small radius (in this experiment, the sphere radius was taken as $1.0$
in arbitrary units). This choice is motivated by the following considerations.
First, starting from a compact configuration (in which all residues
are concentrated near the origin) avoids situations where the chain
is completely extended and the folding process would require excessively
long times to reach equilibrium globular states. Second, we deliberately
exclude from consideration macrostates corresponding to distributions
in the subspace of extended configurations, since they are certainly
not related to the region of phase space of interest to us, associated
with compact structures. Furthermore, within a single sequence, we
set the same initial coordinates for all replicas. This was done to
reduce the probability that different replicas would initially find
themselves in significantly different regions of configuration space.
With the same start, all replicas have equal chances to converge to
the same or different macrostates, but the influence of randomness
of initial conditions is minimized. This approach allows more clearly
identifying effects associated precisely with the shape of the energy
landscape, rather than with trivial scatter of initial configurations.

In the first stage, all 50 sequences were processed. The key results
can be summarized as follows. All 50 replicas in each of the 50 sequences
were found to be unique. Of the 50 sequences, ultrametricity (i.e.,
fraction of ultrametric triangles $f_{ultra}^{(k)}>0.5$) was detected
in 45 sequences, which constitutes $90.0\%$ of the total. Among the
sequences demonstrating ultrametricity, in the vast majority (44 out
of 45, i.e., in $97.8\%$ of cases), nontrivial ultrametricity predominates,
meaning the fraction of triangles satisfying the conditions of isoscelesness
with a clear distinction between the small and medium distances exceeds
the fraction of trivially equilateral triangles. Only in one sequence
was predominance of trivial ultrametricity observed. The average fractions
of triangles over the ensemble of sequences were: trivial $19.31\%\pm9.26\%$,
nontrivial $46.07\%\pm5.24\%$, total ultrametricity $65.38\%\pm12.48\%$.
The high standard deviation for the trivial and total fractions indicates
significant variability of landscapes from sequence to sequence. The
Edwards--Anderson parameter, averaged over the ensemble of sequences,
was $\langle q_{EA}\rangle=0.292\pm0.066$. This value is noticeably
above zero and indicates that the system is near the glass transition
temperature, where ergodicity is broken and many different macrostates
exist. The range of $q_{EA}$ lies between $0.156$ and $0.433$,
confirming that even within a single model, different sequences can
lead to different degrees of "glassiness." The average maximum distance
between residues in the collected configurations was about $7.14$
in arbitrary units, which corresponds to a compact globular structure
(for comparison, a completely extended chain of length 128 with step
$1$ would have a length on the order of $127$). Based on the results
of the first stage, all 44 sequences in which nontrivial ultrametricity
was detected were selected for the second stage.

The second stage was performed for the selected 44 sequences with
a changed random number generator seed, which ensured independence
of the Monte Carlo trajectories from the first stage. All other model
and simulation parameters remained unchanged. Of the 44 sequences,
ultrametricity was confirmed for 42 sequences ($95.5\%$). Two sequences
no longer satisfied the criterion $f_{ultra}>0.5$. In 41 out of 42
cases ($97.7\%$), nontrivial ultrametricity again predominated. The
average fractions of triangles over the ensemble of the second stage
were: trivial $19.93\%\pm8.86\%$, nontrivial $47.67\%\pm3.03\%$,
total ultrametricity $67.59\%\pm9.23\%$. These values are close to
those obtained in the first stage, indicating the statistical stability
of the observed patterns. The average value of $q_{EA}$ was $0.284\pm0.055$,
practically coinciding with the first stage. The average maximum distance
also remained practically at the previous level --- $7.11$. The
fact that two sequences out of 44 no longer met the ultrametricity
criterion upon repeated simulation can be explained by two main reasons.
First, the change in initial initialization and subsequent annealing
could have led to the set of achieved macrostates for this sequence
being different. As a result, among the replicas, those replicas that
formed the ultrametric subset could be absent, or the subset of replicas
itself could have changed. Second, during the Monte Carlo dynamics
associated with collecting statistically independent configurations,
some replicas could over time, with some (though not very large) probability,
change their basin of attraction and end up in a different valley,
which could lead to distortion of the collected statistical data.
The reason for this is that although we use adaptive step size and
simulated annealing, it is impossible to completely exclude transitions
between macrostates in a finite simulation, especially if the energy
barriers are not too high. Thus, the observed "disappearance" of
ultrametricity for some sequences does not contradict the overall
picture but rather emphasizes the sensitivity of the hierarchical
structure to the specific set of replicas generated by stochastic
dynamics.

\begin{figure}[H]
\centering{}\centering \includegraphics[width=0.8\textwidth]{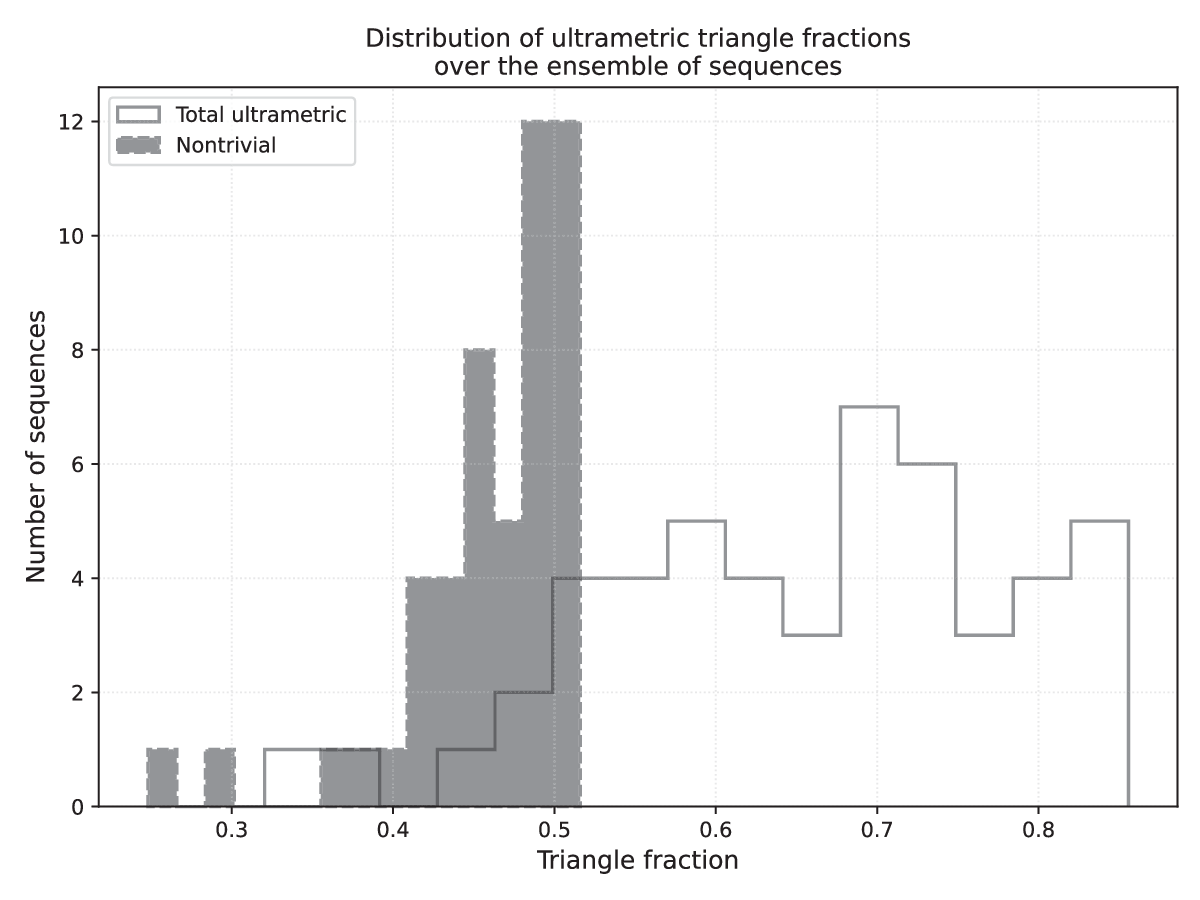}
\caption{Histogram of the distribution of fractions of ultrametric triangles
over the ensemble of 50 random sequences (first stage). Columns outlined
with a solid line show the distribution for the total fraction of
ultrametric triangles $f_{\text{ultra}}$. Columns outlined with a
dashed line with gray fill show the distribution for the fraction
of nontrivial ultrametric triangles $f_{\text{nontriv}}$. The peak
of the $f_{\text{nontriv}}$ distribution occurs in the region $0.45-0.50$,
indicating the predominance of nontrivial hierarchical organization.}
\label{fig:hist}
\end{figure}

\begin{figure}[H]
\centering{}\centering \includegraphics[width=0.8\textwidth]{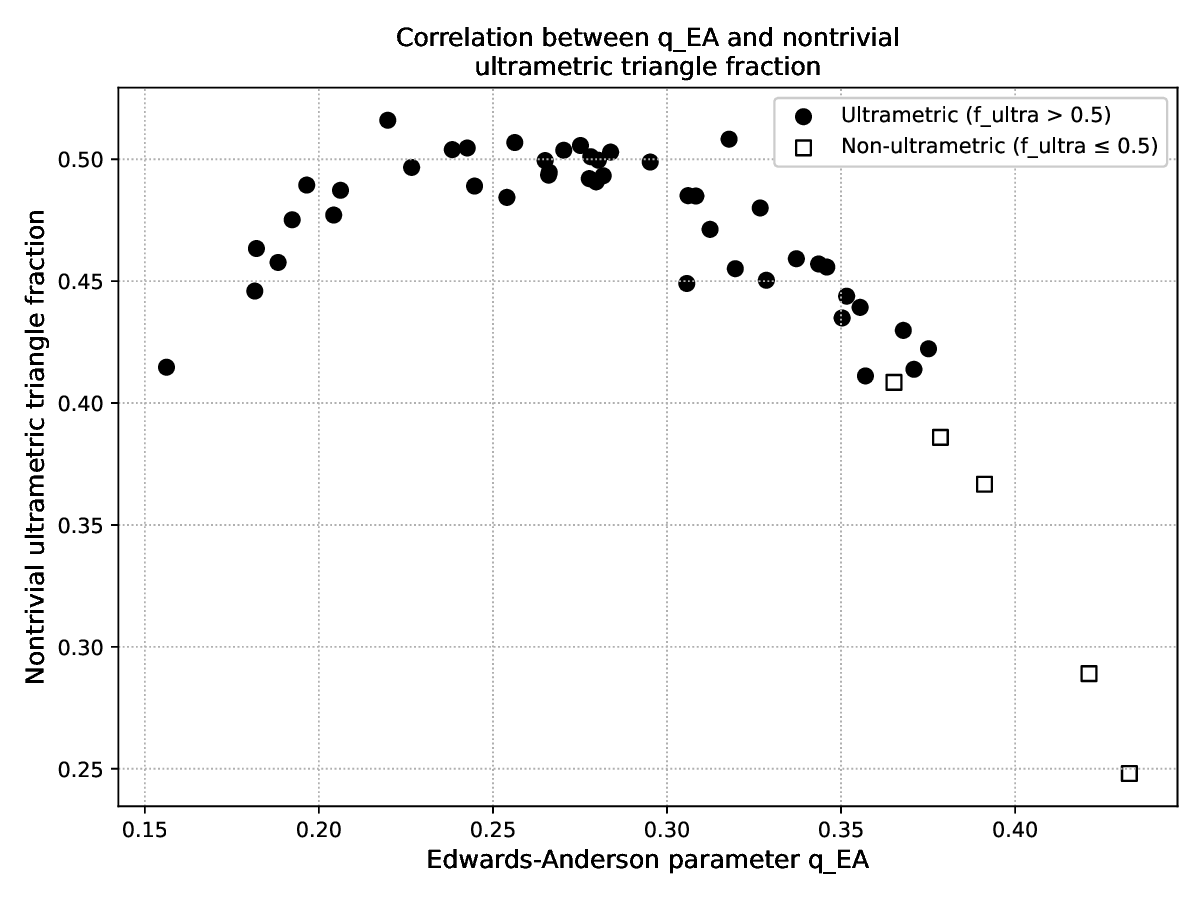}
\caption{Correlation between the Edwards--Anderson parameter $q_{\text{EA}}$
and the fraction of nontrivial ultrametric triangles $f_{\text{nontriv}}$
for 50 sequences (first stage). Filled black circles denote sequences
for which the total fraction of ultrametric triangles exceeds 0.5
(ultrametric), empty squares denote the remaining (non-ultrametric)
ones. A non-monotonic structure is observed: maximum values of $f_{\text{nontriv}}$
are achieved at $q_{\text{EA}}\approx0.25-0.30$, indicating the existence
of an optimal level of frustration for the formation of a hierarchical
landscape. At higher values of $q_{\text{EA}}$ (\textgreater 0.35),
ultrametricity is typically not detected.}
\label{fig:corr}
\end{figure}

\begin{figure}[H]
\centering{}\centering \includegraphics[width=1\textwidth]{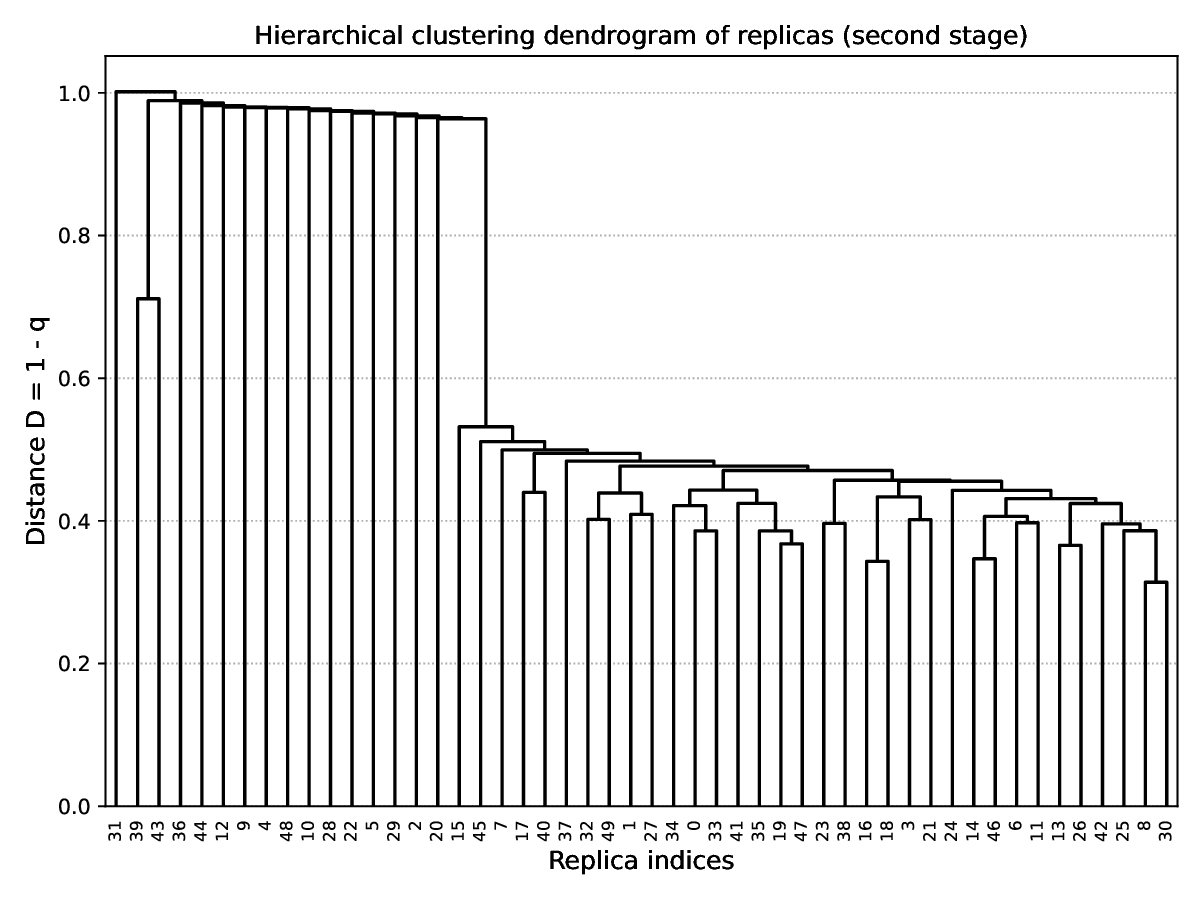}
\caption{Dendrogram of hierarchical clustering of 50 replicas for the sequence
with the maximum fraction of nontrivial ultrametric triangles at the
second stage (sequence with $f_{\text{nontriv}}=51.02\%$, $f_{\text{ultra}}=76.37\%$).
The distance matrix is defined as $D=1-q$, where $q$ is the overlap
between replicas. The average linkage method was used in constructing
the dendrogram. An asymmetric structure is clearly visible: in the
left part of the dendrogram, the merging levels of clusters ($D\sim0.7-0.9$)
are significantly higher than in the right part ($D\sim0.3-0.5$).
This corresponds to a hierarchy of macrostates: large clusters separated
by high barriers (left part) contain internal subclusters corresponding
to closer substates (right part). Such a structure is a characteristic
feature of ultrametric organization of state space.}
\label{fig:dendro}
\end{figure}

The simulation results are presented in Figures~\ref{fig:hist}--\ref{fig:dendro}.

Figure~\ref{fig:hist} shows the distributions of fractions of ultrametric
triangles over the ensemble of 50 random sequences at the first stage.
Columns outlined with solid lines without fill show the distribution
of the total fraction of all ultrametric triangles $f_{\text{ultra}}$,
and columns with gray fill outlined with dashed lines show the distribution
of the fraction of nontrivial ultrametric triangles $f_{\text{nontriv}}$.
Both distributions have distinct maxima: for $f_{\text{ultra}}$ in
the region $0.65$--$0.75$, for $f_{\text{nontriv}}$ in the region
$0.45$--$0.50$. The asymmetry of the $f_{\text{nontriv}}$ distribution
with a gentler left slope indicates that a significant portion of
sequences have moderate values of nontrivial ultrametricity. Note
that $f_{\text{ultra}}=f_{\text{triv}}+f_{\text{nontriv}}$, and from
the first stage data for the averages we have $\langle f_{\text{ultra}}\rangle\sim68\%$,
$\langle f_{\text{nontriv}}\rangle\sim48\%$, $\langle f_{\text{triv}}\rangle\sim20\%$.
Thus, the contribution of the trivial component is not dominant but
is also not negligible. The fact that a significant part of sequences
have $f_{\text{nontriv}}\sim0.5$ should not cause surprise. Nontrivial
ultrametricity is a stricter criterion of hierarchical organization
than total ultrametricity. In real systems with finite fluctuations,
there will always be some number of "borderline" triangles that
are classified as trivially ultrametric. In this case, in the vast
majority of ultrametric sequences, predominance of the nontrivial
component is observed ($f_{\text{nontriv}}>f_{\text{triv}}$).

Figure~\ref{fig:corr} presents a graph of the correlation between
the Edwards--Anderson parameter $q_{\text{EA}}$ and the fraction
of nontrivial ultrametric triangles $f_{\text{nontriv}}$ for 50 sequences
at the first stage. Filled black circles denote sequences for which
the total fraction of ultrametric triangles exceeds $0.5$ (ultrametric),
empty squares denote the remaining (non-ultrametric) sequences. A
non-monotonic structure of the point distribution is clearly observed,
with a maximum in the region $q_{\text{EA}}\approx0.2-0.30$, where
$f_{\text{nontriv}}$ reaches values $\sim0.50$. This indicates the
existence of an optimal level of frustration at which the hierarchical
organization of the energy landscape is most pronounced. At lower
values of $q_{\text{EA}}$, the system is insufficiently frustrated
to form a developed hierarchy, while at high $q_{\text{EA}}$ ($>0.35$),
excessive frustration leads to "blurring" of boundaries between
macrostates and suppression of ultrametricity. Sequences not satisfying
the ultrametricity criterion are concentrated predominantly in the
region of high $q_{\text{EA}}>0.35$, confirming the connection between
strong frustration and loss of hierarchical structure.

Figure~\ref{fig:dendro} shows a dendrogram of hierarchical clustering
of 50 replicas for the sequence with the maximum fraction of nontrivial
ultrametric triangles at the second stage (sequence with $f_{\text{nontriv}}=51.02\%$,
$f_{\text{ultra}}=76.37\%$). The distance matrix is defined as $D=1-q$,
where $q$ is the overlap between replicas. The average linkage method
was used in constructing the dendrogram. Characteristic asymmetry
is clearly visible on the graph: the merging levels of clusters in
the left part of the dendrogram are significantly higher than in the
right part. High merging levels in the left part ($D\sim0.75\div0.95$)
correspond to the merging of large clusters of macrostates separated
by significant energy barriers. These clusters represent fundamentally
different conformational families (e.g., chain folds with different
spatial architectures). Transition between such families requires
overcoming high barriers and is kinetically forbidden under low-temperature
conditions, which is reflected in the large distance between the corresponding
replicas. In the right part of the dendrogram, merging levels are
located significantly lower ($D\approx0.3-0.5$). This indicates the
existence of a hierarchy of substates within large clusters. Replicas
are grouped into compact groups at small distances, then these groups
merge at intermediate distances, and only then merging with other
large clusters occurs at high levels. It is important to note that
such asymmetry (high levels on the left, low on the right) is not
an artifact of a specific sequence but reflects a general property
of ultrametric spaces: the existence of a distinguished "root" of
the hierarchy and many "leaves" grouping at different levels. In
terms of spin glass theory, this corresponds to the structure of the
Parisi tree, where pure states are organized into a hierarchy with
different scales of overlaps. The observed dendrogram demonstrates
that even in our simplified protein prototype model, the space of
macrostates exhibits a similar structure.

Analysis of the obtained results allows drawing several important
conclusions. First of all, even in such a simplified model, where
the protein is represented only by a chain of point particles with
a limited set of interactions and with heuristic selection of potential
parameters and classification thresholds, the energy landscape exhibits
clear signs of ultrametric organization. This is manifested in the
fact that, according to the results of two stages, for the vast majority
of random sequences ($80\%$), the distance matrix between replicas
contains more than $50\%$ of triangles satisfying the ultrametric
inequality, in which the nontrivial type of ultrametricity predominates.
This points to a real hierarchical structure, rather than an artifact
of equilateral triangles that could arise, for example, from random
scatter of points in a high-dimensional space. It is important to
emphasize that ultrametricity is not required to manifest in all sequences
without exception. The structure of the energy landscape is determined
by the specific set of interactions encoded by the amino acid sequence.
In some disorder realizations, frustrations may be weak, and the system
may behave more like a simple glass without a pronounced hierarchy,
or like a system with a single macrostate. Our results show that about
$20\%$ of sequences do not yield a sufficient fraction of ultrametric
triangles ($f_{ultra}\le0.5$). This is natural and expected. Moreover,
even within a single sequence possessing an ultrametric landscape,
not all possible triples of replicas are required to be ultrametric.
Ultrametricity is a property that may hold for some subset of states
corresponding, for example, to those macrostates responsible for globular
structures with a certain architecture. In our simulation, we observe
that the fraction of ultrametric triangles ranges from $40\%$ to
$87\%$, indicating that the hierarchy covers a significant, but not
necessarily all, set of macrostates.

Thus, the totality of obtained data --- the high fraction of nontrivially
ultrametric sequences, the stability of this property upon repeated
simulation, the presence of an optimal level of frustration, and the
characteristic asymmetry of dendrograms --- indicates that the heuristically
chosen potential parameters and classification thresholds indeed allowed
revealing a property intrinsic to the model: hierarchical organization.

\section{Conclusion and perspectives}

In this work, we have presented and implemented a computational scheme
for testing the ultrametricity of the energy landscape in a simple
model of a protein prototype, where amino acid residues are modeled
as material points, and pairwise interactions include universal repulsion,
the Lennard-Jones potential for hydrophobic residues, Coulomb interaction
with screening for charged residues, and elastic bonds between neighbors.
A key feature of this scheme is the definition of macroscopic overlap
between replicas through the correlation of vectors of mean pairwise
energies, which is completely analogous to the approach in spin glass
theory and allows applying the statistical apparatus of replica theory.
It is important to emphasize that, in contrast to the classical approach
in the theory of disordered heteropolymers, in this work we fundamentally
refrain from the procedure of averaging over disorder. The analysis
is performed for each specific residue sequence independently, which
corresponds to the uniqueness of each molecule and allows revealing
the connection between a specific disorder realization and the structure
of its energy landscape. The computational algorithm includes stages
of generating an ensemble of random sequences, achieving equilibrium
with simulated annealing and adaptive Monte Carlo step size, collecting
statistically independent configurations, computing overlap matrices,
and analyzing ultrametricity with discrimination between trivial and
nontrivial types.

Numerical modeling was performed for a system with chain length $N=128$,
number of sequences $K=50$, number of replicas $M=50$, and sample
size $S=400$ for each replica. The results obtained indicate that
even in an extremely simplified model, the energy landscape exhibits
pronounced signs of ultrametric organization. In the vast majority
of random sequences, a high fraction of ultrametric triangles was
recorded in the distance matrix between replicas, with the dominant
contribution coming from the nontrivial component. This points to
the presence of a hierarchical structure, rather than an artifact
of equilateral triangles characteristic of trivial ultrametricity.
The values of the Edwards--Anderson parameter confirm that the system
is near the spin glass phase with broken ergodicity. A repeated experiment
for selected sequences demonstrated the statistical stability of the
observations, although a small fraction of systems lost their ultrametric
property, which is explained by the stochastic nature of the Monte
Carlo method and possible mixing of replicas between macrostates.
Correlation analysis revealed a non-monotonic dependence between the
degree of frustration and the level of ultrametricity, indicating
the existence of an optimal level of disorder favorable for the formation
of a hierarchical landscape. Dendrograms for sequences with maximal
ultrametricity clearly demonstrate an asymmetric structure: large
clusters corresponding to fundamentally different conformational families
are separated by large distances, while within them a finer hierarchy
of substates is observed. These results provide additional numerical
arguments in favor of Frauenfelder's hypothesis about the hierarchy
of conformational substates of proteins \cite{frauenfelder1988,frauenfelder1991,frauenfelder1999}
and show that ultrametricity may be an intrinsic property of even
very coarse models that imitate only the most general features of
polypeptide chains. The high fraction of sequences with nontrivial
ultrametricity, obtained in the absence of any angular interactions
or specific orientational effects, suggests that hierarchical organization
is a fundamental consequence of the competition between long-range
and short-range forces in a polymer chain with disorder.

In our opinion, further research in this direction should be aimed
at a systematic study of how the complication of the model affects
the ultrametric properties of the energy landscape. To this end, one
should sequentially introduce new elements into the model, starting
from the simplest and gradually moving to more complex descriptions.
At each stage, it is necessary to verify whether ultrametricity is
preserved and how its quantitative characteristics change. Such an
approach will allow identifying the minimal set of physical interactions
necessary for the formation of hierarchical structure.

First of all, the dependence of ultrametricity on temperature should
be studied. It is expected that with decreasing temperature, the degree
of ultrametricity will increase, reaching saturation in the spin glass
phase, whereas at high temperatures the landscape should become simpler,
more ergodic, which should lead to a decrease in the fraction of nontrivial
ultrametric triangles. Such behavior is predicted by spin glass theory
and has been observed in numerical experiments for the Edwards--Anderson
model \cite{ogielski1985,marinari1998,contucci2007}. It is important
to quantitatively characterize this transition by determining the
glass transition temperature $T_{g}$ for this model through the temperature
dependence of the Edwards--Anderson parameter. Next, the dependence
of the degree of ultrametricity on system size $N$ should be investigated,
since with increasing chain length, the number of degrees of freedom
increases, and one can expect a complication of the hierarchical structure.
Also important is the systematic variation of sequence composition
(probabilities of occurrence of hydrophobic, charged, and neutral
residues). This will allow constructing a phase diagram in composition
space and identifying regions where ultrametricity is most pronounced.

Despite the obtained evidence in favor of ultrametricity, it is necessary
to be aware of the limitations of the model. These limitations may
affect the generality of our conclusions and must be overcome when
moving towards more realistic descriptions. The main simplification
is the representation of amino acid residues as material points devoid
of spatial structure. In reality, each residue has volume, shape,
and a specific orientation in space, which is critically important
for the specificity of interactions. Neglecting these factors means
that we completely ignore steric effects (related to side chain packing)
and the possibilities of forming directed interactions, such as hydrogen
bonds and ionic pair potentials, which require a specific mutual orientation
of partners. Furthermore, our model lacks any angular degrees of freedom:
a real polypeptide chain possesses rigidity associated with valence
and torsion angles, which leads to the formation of secondary structure
elements --- alpha helices and beta sheets. These structures are
not only stabilized by hydrogen bonds but also impose strong constraints
on the configuration space, potentially creating additional levels
of hierarchy. The absence of explicit solvent modeling also limits
the realism of the model: hydrophobic attraction is accounted for
only through an effective pairwise Lennard-Jones potential, which
does not capture the collective nature of the hydrophobic effect,
and also completely ignores medium viscosity and hydration dynamics.
In addition, the potentials used in the model are extremely coarse:
the Lennard-Jones potential with uniform parameters for all hydrophobic
residues does not reflect differences between, for example, aliphatic
and aromatic side chains, and Coulomb interaction with constant screening
does not account for the inhomogeneity of the dielectric constant
inside the protein and near its surface.

The enumerated limitations naturally indicate ways to extend the model,
which will bring it closer to real proteins and allow testing the
stability of ultrametric properties under more realistic conditions.
The first and most conceptually important step is to abandon the point
particle approximation and introduce the spatial structure of residues.
In the original model, each residue is represented by a single point
in three-dimensional space. The generalization consists in modeling
each residue $i$ by a set of $L$ interacting centers (sites) $\{r_{i,1},r_{i,2},\dots,r_{i,L}\}$,
where $r_{i,\alpha}\in\mathbb{R}^{3}$ are the coordinates of the
$\alpha$-th site of residue $i$. These sites are rigidly connected
to each other, meaning their mutual arrangement is fixed and determines
the shape of the residue. The state of residue $i$ as a rigid body
is completely described by six degrees of freedom. To specify them,
we choose some center in the residue (e.g., the center of mass or
the position of the $C_{\alpha}$ atom) with coordinates $R_{i}\in\mathbb{R}^{3}$.
We specify the orientation of the residue by a right-handed triple
of orthonormal vectors $u_{i},v_{i},w_{i}\in\mathbb{R}^{3}$. These
vectors define the orientation of the local coordinate system associated
with the residue relative to the laboratory system. The internal geometry
of the residue is determined by the residue type, which we denote
by $\tau$. For each type $\tau$, the coordinates of all its sites
in the local coordinate system are given. Denote for a residue of
type $\tau$ the coordinates of its $\alpha$-th site as a vector:
\[
a_{\tau,\alpha}=(a_{\tau,\alpha}^{x},a_{\tau,\alpha}^{y},a_{\tau,\alpha}^{z})\in\mathbb{R}^{3},
\]
where $\alpha=1,\dots,L_{\tau}$, and $L_{\tau}$ is the number of
sites for residues of this type. Each vector $a_{\tau,\alpha}$ is
a model parameter and does not change during the evolution of the
system. The coordinates of the $\alpha$-th site of residue $i$ (of
type $\tau_{i}$) in the laboratory coordinate system are expressed
through the dynamic variables $R_{i},u_{i},v_{i},w_{i}$ and model
parameters $a_{\tau_{i},\alpha}$ as follows:
\begin{equation}
r_{i,\alpha}=R_{i}+a_{\tau_{i},\alpha}^{x}u_{i}+a_{\tau_{i},\alpha}^{y}v_{i}+a_{\tau_{i},\alpha}^{z}w_{i}.\label{r_i_alpha}
\end{equation}
The interaction energy between two residues $i$ and $j$ (of types
$\tau_{i}$ and $\tau_{j}$, respectively) equals the sum of pairwise
interactions between all their sites:
\[
U_{ij}\left(R_{i},u_{i},v_{i},w_{i},R_{j},u_{j},v_{j},w_{j}\right)=\sum_{\alpha=1}^{L_{\tau_{i}}}\sum_{\beta=1}^{L_{\tau_{j}}}U_{\alpha\beta}\left(r_{i,\alpha},r_{j,\beta}\right),\tag{2}
\]
where $r_{i,\alpha}$ and $r_{j,\beta}$ are computed by formula (\ref{r_i_alpha}),
and $U_{\alpha\beta}$ is the interaction potential between sites
$\alpha$ and $\beta$. It is important to emphasize that the potential
$U_{\alpha\beta}$ depends only on the coordinates of the sites. The
dependence on orientation arises automatically, since the site coordinates,
being arguments of $U_{\alpha\beta}$, themselves depend on orientation
through formula (\ref{r_i_alpha}).

In the simplest case, the interaction between sites may depend only
on the distance between them. Then
\[
U_{\alpha\beta}\left(r_{i,\alpha},r_{j,\beta}\right)=U_{\alpha\beta}\left(|r_{i,\alpha}-r_{j,\beta}|\right).
\]
To model attraction between specific types of sites, one can use the
Lennard-Jones potential of the form (\ref{L_D}). For charged sites,
one can add Coulomb interaction with screening of the form (\ref{Coul}).
To account for more subtle effects related to the mutual orientation
of sites (for example, if the sites represent planar groups), the
potential may depend not only on distance but also on direction. In
this case, potentials can be used whose arguments include vectors
constructed from site coordinates. For example, if site $\alpha$
has a distinguished direction given by a unit vector $n_{i,\alpha}$,
which is also rigidly connected to the residue and computed through
$u_{i},v_{i},w_{i}$ similarly to coordinates, then one can define
a potential of the form:
\[
U_{\text{orient}}(r_{i,\alpha},r_{j,\beta},n_{i,\alpha},n_{j,\beta})=\varepsilon_{\text{orient}}\left[\left(\frac{\sigma_{\text{orient}}}{|r_{i,\alpha}-r_{j,\beta}|}\right)^{12}-\left(\frac{\sigma_{\text{orient}}}{|r_{i,\alpha}-r_{j,\beta}|}\right)^{6}\right]\cdot f(n_{i,\alpha},n_{j,\beta},r_{i,\alpha}-r_{j,\beta}),
\]
where the function $f$ depends on scalar products of the type $n_{i,\alpha}\cdot n_{j,\beta}$
and $n_{i,\alpha}\cdot(r_{i,\alpha}-r_{j,\beta})$. Such a functional
form allows modeling preferred mutual orientations without reference
to specific chemistry.

Potentials defining chain geometry can be implemented as follows.
In real polymer chains, neighboring monomers are connected by covalent
bonds, which imposes constraints on the distances between them, as
well as on the angles between bonds. To account for these constraints
in the model, one can introduce potentials depending on the mutual
arrangement of consecutive residues along the chain. If a distinguished
center is defined for each residue $i$ (e.g., the center of mass
or the position of the $C_{\alpha}$ atom) with coordinates $R_{i}$,
then the distance between centers of neighboring residues $|R_{i+1}-R_{i}|$
can be maintained near a fixed value using an elastic potential of
the form (\ref{Bond}). To describe valence angles formed by three
consecutive residues, one can use a potential of the form
\begin{equation}
U_{\text{angle}}(\theta_{i})=\frac{1}{2}k_{\theta}(\theta_{i}-\theta_{0})^{2},\label{U_angle}
\end{equation}
where $\theta_{i}$ is the angle between the vectors $R_{i}-R_{i-1}$
and $R_{i+1}-R_{i}$, the parameter $k_{\theta}>0$ is the angular
stiffness, and the parameter $\theta_{0}$ is the equilibrium value
of the valence angle. To describe torsion angles, which determine
rotation around bonds, one can introduce a potential depending on
the dihedral angle $\phi_{i}$ formed by four residues $i-2,i-1,i,i+1$:
\begin{equation}
U_{\text{dihedral}}(\phi_{i})=\sum_{n=1}^{N}K_{n}[1+\cos(n\phi_{i}-\delta_{n})],\label{U_dihedral}
\end{equation}
where $K_{n}$ are amplitude parameters, and $\delta_{n}$ are phase
shift parameters. The choice of parameters for these potentials allows
creating preferred chain conformations, for example, favoring helical
or sheet structures.

The potential (\ref{U_angle}) is three-body, and the potential (\ref{U_dihedral})
is four-body. The introduction of many-body potentials requires a
corresponding generalization of the replica comparison method. The
vector $\bar{u}^{(k,m)}$, which we use to define overlap, can no
longer be limited to only pairwise energies, since it would not contain
information about many-body contributions that may be significant
for the structure of the energy landscape. The most natural generalization
consists in including all types of energies present in the model in
this vector. Define for each replica $m$ of sequence $k$ a vector
$\bar{w}^{(k,m)}$, whose components are: average values of pairwise
energies $\langle U_{ij}\rangle$ for all $i<j$; average values of
angular energies $\langle U_{\text{angle}}(\theta_{i})\rangle$ for
all $i$; average values of torsion energies $\langle U_{\text{dihedral}}(\phi_{i})\rangle$
for all $i$. The vector $\bar{w}^{(k,m)}$ has dimension equal to
the total number of all independent energy contributions in the model
and contains all available information about the thermodynamic averages
of the various energy components of the system. Then the overlap between
replicas $m$ and $n$ for sequence $k$ is defined as the Pearson
correlation coefficient according to formula (\ref{corr}) with the
substitution $\bar{u}^{(k,m)}\rightarrow\bar{w}^{(k,m)}$. With this
approach, many-body interactions are naturally accounted for in the
definition of overlap. This allows applying a unified methodology
to models with arbitrary levels of complexity, from the simplest models
with pairwise interactions to models with many-body potentials of
any nature.

Also a natural generalization of the model is the introduction of
a more detailed classification of residues, differing, for example,
in size, shape, or nature of interaction. The original model uses
only four types of residues, differing only in charge and hydrophobicity
indicator. For a more refined study of the influence of sequence composition
on ultrametricity, one can introduce a more detailed classification.
Namely, each residue type $\tau$ can be characterized by the following
set of parameters: number of sites $L_{\tau}$, their relative coordinates
$\{a_{\tau,\alpha}\}_{\alpha=1}^{L_{\tau}}$, site charges $\{q_{\tau,\alpha}\}$,
Lennard-Jones potential parameters $\{\varepsilon_{\text{LJ},\tau,\alpha\beta},\sigma_{\text{LJ},\tau,\alpha\beta}\}$
for interaction between sites of different types, orientational interaction
parameters, etc.

All the described extensions of the model can be introduced sequentially,
which allows re-examining the ultrametric properties of the energy
landscape at each stage. It is important to emphasize that the proposed
analysis methodology --- using macroscopic overlap based on vectors
of mean energies and verifying the ultrametricity of the distance
matrix --- remains fully applicable at any level of model complexity.
In the general case, the vector $\bar{w}^{(k,m)}$ will include all
types of energy contributions: pairwise, three-body, four-body, and
so on. For each configuration, the corresponding values are computed,
which are then averaged over the replica, after which the obtained
vectors are used to calculate overlaps. This approach opens the possibility
for a systematic study of the evolution of ultrametric properties
as the model becomes more complex. The results obtained on the current,
extremely simple model serve as a natural starting point for such
an investigation. The observed picture suggests that the hierarchical
structure of the landscape is embedded already at the most basic level
of statistical description of a heteropolymer with competing interactions,
and subsequent complication of the model merely builds upon this structure.

\end{document}